\documentclass[aip,
 amsmath,amssymb,
 reprint,%
]{revtex4-1}

\usepackage{graphicx}
\usepackage{dcolumn}
\usepackage{bm}
\usepackage{color}

\usepackage[utf8]{inputenc}
\usepackage[T1]{fontenc}
\usepackage{mathptmx}
\usepackage{etoolbox}

\makeatletter
\def\@email#1#2{%
 \endgroup
 \patchcmd{\titleblock@produce}
  {\frontmatter@RRAPformat}
  {\frontmatter@RRAPformat{\produce@RRAP{*#1\href{mailto:#2}{#2}}}\frontmatter@RRAPformat}
  {}{}
}%
\makeatother
\begin{document}

\preprint{AIP/123-QED}

\title[Nucleation and phase transition of decagonal quasicrystals]{Nucleation and phase transition of decagonal quasicrystals}
\author{Tiejun Zhou}
\affiliation{Hunan Key Laboratory for Computation and Simulation in Science and Engineering, Key Laboratory of Intelligent Computing and Information Processing of Ministry of Education, School of Mathematics and Computational Science, Xiangtan University, Xiangtan, Hunan, China, 411105.}

\author{Lei Zhang}%
\email{zhangl@math.pku.edu.cn}
\affiliation{Beijing International Center
for Mathematical Research, Peking University, Beijing, 100871, China.
}%

\author{Pingwen Zhang}%
\email{pzhang@pku.edu.cn}
\affiliation{School of Mathematics and Statistics, Wuhan University, Wuhan, 430072, China.}%
\affiliation{School of Mathematical Sciences, Peking University, Beijing, 100871, China.}

\author{An-Chang Shi}%
\email{shi@mcmaster.ca}
\affiliation{Department of Physics and Astronomy, McMaster University, Hamilton, Canada, L8S 4M1.
}%

\author{Kai Jiang}
\email{kaijiang@xtu.edu.cn}
\affiliation{Hunan Key Laboratory for Computation and Simulation in Science and Engineering, Key Laboratory of Intelligent Computing and Information Processing of Ministry of Education, School of Mathematics and Computational Science, Xiangtan University, Xiangtan, Hunan, China, 411105.}

\date{\today}

\begin{abstract}
In this work, we study the nucleation of quasicrystals from liquid or periodic crystals by developing an efficient order-order phase transition algorithm, namely the nullspace-preserving saddle search method. Specifically, we focus on nucleation and phase transitions of the decagonal quasicrystal (DQC) based on the Lifshitz-Petrich model. We present the nucleation path of DQC from the liquid and demonstrate one- and two-stage transition paths between DQC and periodic crystals. We provide a perspective of the group-subgroup phase transition and nucleation rates to understand the nucleation and phase transition mechanisms involving DQC. These results reveal the one-step and stepwise modes of symmetry breaking or recovery in the phase transition from DQC, where the stepwise modes are more probable.
\end{abstract}

\maketitle
\section{Introduction}
Quasicrystals are space-filling ordered structures possessing rotational symmetry but without translational invariance. They have been a topic of significant interest in the fields of materials science and condensed matter physics since their first discovery\,\cite{shechtman1984metallic}. Numerous quasicrystals with 5-, 8-, 10-, 12-, and 18-fold rotational symmetries have been reported in metallic alloys\,\cite{levine1986quasicrystals} and soft matter\,\cite{zeng2004supramolecular,hayashida2007polymeric,xiao2012dodecagonal}. Moreover, quasicrystals, especially decagonal quasicrystals (DQC), have been studied from various perspectives, including formation\,\cite{fung1986icosahedrally,steurer2005structural,suck2013quasicrystals}, geometric features\,\cite{divincenzo1991quasicrystals,jaric2012introduction}, thermodynamic stability\,\cite{denton1998stability,lifshitz1997theoretical,subramanian2016three}. However, the nucleation mechanism of quasicrystals from the liquid and periodic crystals is a topic that requires close attention\,\cite{ash2022other}.

Over the past decades, much attention has been devoted to studying the nucleation and phase transition of quasicrystals since the first discovery, but little progress has been gained. Experimentally, only some snapshots of nucleation phenomena 
have been observed because the nucleation events are rare and occur at ultrahigh speeds\,\cite{reyes1992nucleation,smerdon2008nucleation,hornfeck2014quasicrystal,kurtuldu2018metastable,wang2024al}. In particular, the critical nuclei of quasicrystals usually possess very high surface energies and are extremely unstable, existing for very short time. It is difficult to capture the critical nuclei and the kinetic process of quasicrystal nucleation due to the lack of precise characterization techniques. Theoretical studies could provide an effective way to study nucleation and phase transition. Several mainstream theoretical frameworks include atomistic simulations\,\cite{kashchiev2000nucleation}, density functional theories\,\cite{cahn1959free} and Landau theories\,\cite{cowley1980structural}. These theories have been successfully applied to study nucleation and phase transition of the liquid, periodic crystals, liquid crystals, etc\,\cite{cheng2010nucleation,lin2010numerical,li2013nucleation,samanta2014microscopic}. However, 
in the case of quasicrystals, previous works have focused on studying the growth of quasicrystals by imposing initial embryos\,\cite{achim2014growth,schmiedeberg2017dislocation,tang2020atomic,jiang2020growth,liang2022molecular}. The spontaneous nucleation of quasicrystals has rarely been reported and remains a challenge in simulation, limited by the computational methods to locate critical nuclei.

Theoretically, the key to studying nucleation is to find the index-1 saddle points (transition states, critical nuclei) on the free energy landscape. Ordered phases, such as quasicrystals and periodic crystals, usually correspond to degenerate local minima whose Hessian has a nullspace. When searching for a transition state from a degenerate local minimum, the presence of nullspaces makes escaping from the basin very difficult. Furthermore, since quasicrystals and crystals are incommensurate, there are no obvious epitaxial relations
between them. How to represent quasicrystals and crystals in the same computational framework also increases the difficulties in theory. Thus, designing suitable methods for degenerate saddle search problems has long been an important issue. In recent years, two effective saddle search methods for degenerate saddle search problems have been proposed, including the high-index saddle dynamics method\,\cite{yin2019high,yin2021transition} and the nullspace-preserving saddle search (NPSS) method\,\cite{cui2024efficient}. The former escapes the basin by climbing along an ascent space that includes the nullspace and an ascent direction, while the latter climbs upward along an ascent direction orthogonal to the nullspace of the initial states. These works have offered an efficient methods to study the nucleations and phase transitions of the quasicrystals. Based on these two methods, the transition paths of the dodecagonal quasicrystal have been revealed\,\cite{yin2021transition, cui2024efficient}. However, for the DQC, its nucleation and transition mechanisms are still unclear. For example, what structures can undergo phase transitions to or from DQC? What intermediate states are involved? How does the symmetry transform?

In this work, we focus on the study of nucleation and phase transitions involving DQC based on the Lifshitz-Petrich (LP) model. In particular, we identify three new stable phases based on the symmetry of DQC, including the fusiform crystal (FC), pseudo-6-fold crystal (PC6), and lamellar quasicrystal (LQ). Using the NPSS method, we obtain transition states and minimum energy paths (MEPs) representing the most probable transition paths between
different stable phases. We show that the nucleation path of DQC from the liquid, and demonstrate one- and two-stage transition paths between DQC and crystals. We analyze these phase transitions with the group-subgroup phase transition theory\,\cite{gernoth2001analytical} and nucleation rates. These results reveal that the phase transition from DQC could follow one-step and stepwise modes of symmetry breaking or recovery, where the stepwise modes of breaking or recovery are easier to occur.

\section{Theoretical framework}
\subsection{The Lifshitz-Petrich model}
The LP model provides a useful framework for characterizing the phases and phase transitions of quasiperiodic systems, including the bifrequency excited Faraday wave\,\cite{lifshitz1997theoretical} and soft-matter quasicrystals\,\cite{lifshitz2007soft,jiang2015stability}. The LP model introduces a scalar order parameter $\psi(\bm{r})$ to represent the density profile in a domain $\Omega$. The LP free energy is
\begin{equation}
\mathcal{F}_{LP}(\psi)= \int_{\Omega} \Big\{\frac{1}{2}[(1^2+\Delta)(q^2 +\Delta)\psi]^{2}-\frac{\tau}{2} \psi^{2}-\frac{\gamma}{3} \psi^{3}+\frac{1}{4} \psi^{4}\Big\}\, d\bm{r} \label{eq:LP}
\end{equation}
where $\tau$ is a temperature-like controlling parameter
and $\gamma$ characterizes the intensity of three-body interaction\,\cite{barkan2011stability,jiang2015stability}. $1$ and $q$ are two characteristic length scales, which needed to stabilize the quasicrystals. Furthermore, we impose the mean zero condition of order parameter on the LP systems to ensure the mass conservation $\int_{\Omega} \psi(\bm{r})\,d\bm{r} = 0$,
which comes from the definition of the order parameter, i.e., the deviation from average density. 

\begin{figure*}[!htpb]
	\centering
	\includegraphics[width=1.01\textwidth]{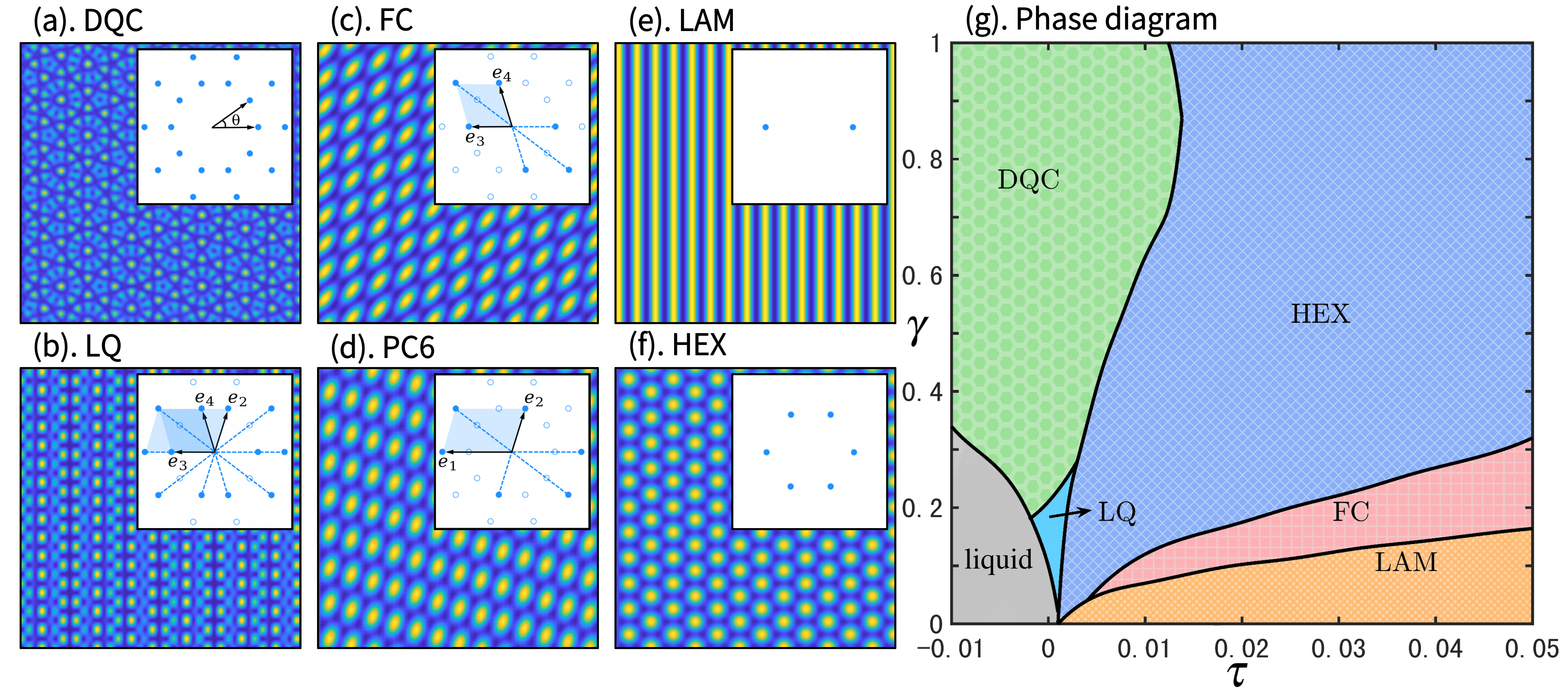}
	\caption{Stable ordered structures (a-f) and phase diagram (g) of the LP model with $q =2\cos(\pi/5)$. (a).\,DQC; (b).\,LQ; (c).\,FC; (d).\,PC6; (e).\,The lamella (LAM) phase; (f).\,The hexagonal (HEX) phase. (b-d) are new structures associated with 10-fold symmetry, where the hollow points represent the corresponding positions of major diffraction spectra of DQC, and the solid points are the major diffraction spectra of new structures.}
	\label{fig:structures_and_phasediagram}
\end{figure*}

In this work, we study the nucleation and phase transition of two-dimensional DQC based on the LP model. To stabilize the DQC, the second length scale $q$ is selected as $2\cos(\pi/5)$ as dictated by the 10-fold symmetry\,\cite{jiang2015stability}. FIG.\,\ref{fig:structures_and_phasediagram}\,(a-d) present four stable structures associated with 10-fold symmetry in physical and reciprocal spaces, including DQC, PC6, FC, and LQ phases. FC and PC6 are new periodic crystals obtained by locating a set of major diffraction spectra forming a parallelogram with the central point in the reciprocal space. Their reciprocal lattice vectors (RLVs) are linear combinations of basic vectors $\{\bm{e}_1,\bm{e}_2\}$ and $\{\bm{e}_3,\bm{e}_4\}$ with integer coefficients respectively, as shown in FIG.\,\ref{fig:structures_and_phasediagram}\,(c) and (d). LQ is a new one-dimensional quasicrystal identified by spanning the basic vectors of PC6 and FC, where $\{\bm{e}_2,\bm{e}_3,\bm{e}_4\}$ is the linearly independent basis over the rational number field. 

We construct the phase diagram in the $\tau-\gamma$ plane using an open software AGPD\,\cite{jiang2022automatically},
which can quickly search for stable states using efficient numerical methods, such as the adaptive accelerated Bregman proximal gradient methods\,\cite{jiang2020efficient,bao2024convergence}. We give converged ordered structures and their initial configurations in the reciprocal space, as shown in FIG.\,\ref{fig:structures_and_phasediagram}\,(a-f). FIG.\,\ref{fig:structures_and_phasediagram}\,(g) presents the phase diagram of the LP model with $q = 2\cos(\pi/5)$.  

\subsection{Incommensurate epitaxies}
\begin{figure}[!hbpt]
	\centering
	\includegraphics[width=0.3\textwidth]{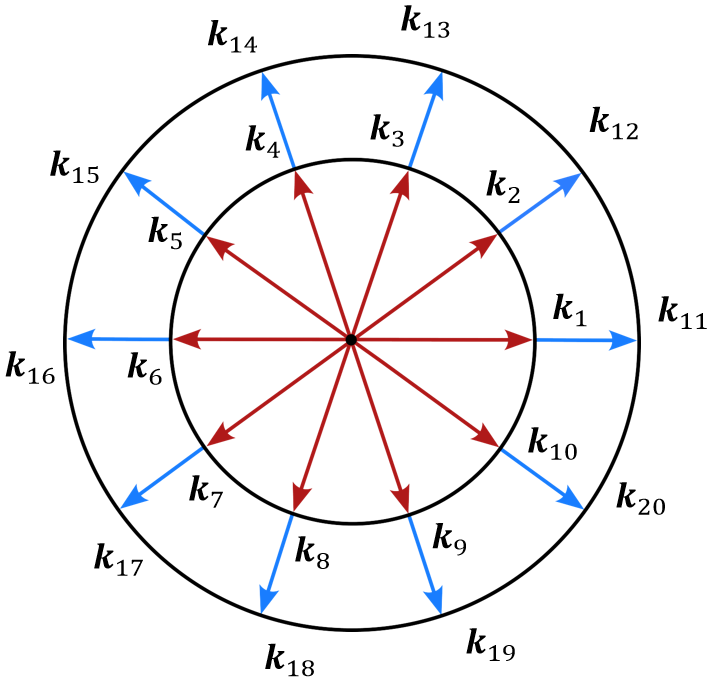}
	\caption{RLVs of two-dimensional DQC, where $\bm{k}_j$ = ($\cos(j\pi/5)$, $\sin(j\pi/5)$) for $j$ = 1, 2, ..., 10, and
    $\bm{k}_j$ = ($q\cos(j\pi/5)$, $q\sin(j\pi/5)$) for $j$ = 11, 12, ..., 20.}
	\label{fig:DQC_reciprocal}
\end{figure}

 One bottleneck in studying phase transitions is that quasicrystals and crystals are incommensurate, i.e., their lattice mismatch\,\cite{janssen2006incommensurate}. In the case of the two-dimensional DQC whose major RLVs are shown in FIG.\,\ref{fig:DQC_reciprocal}, it is impossible to represent all its RLVs linearly using two noncollinear bases over the rational number field.

Methods to solve the lattice mismatch problem between crystals and quasicrystals include the projection method\,\cite{jiang2014numerical} and the periodic approximation method\,\cite{zhang2008efficient}. The former simulates the quasiperiodic structure in a higher-dimensional periodic system, while the latter approximates the quasiperiodic structure with a periodic approximant.
In this paper, we use the periodic approximation method to study phase transitions about DQC. In the computation, some RLVs are linear combinations of primitive RLVs with irrational coefficients, which are difficult for the current computers to store. For the irrational number $\kappa$, we can approximate it with the rational number $[L\kappa]$, where $[a]$ rounds $a$ to the nearest integer. Applying Diophantine approximation theory, the proper value $L$ satisfying the required accuracy can be determined by the Diophantine approximation error (DAE)\,\cite{davenport1946simultaneous}. Then the quasiperiodic structure can be approximated by the periodic structure in a finite domain with the period $[0, 2\pi L)^2$.
For two primitive RLVs $\bm{e}^*_1 = (1, 0)$ and $\bm{e}^*_2 = (0, 1)$, the coefficients of RLVs about DQC required to be approximated simultaneously are listed in Table\,\ref{tbl:Coefficients}.
 
\begin{table}[!hbpt]
\normalsize
  \caption{Coefficients of RLVs for DQC required to be approximated simultaneously.}
  \label{tbl:Coefficients}
  \begin{tabular*}{0.48\textwidth}{@{\extracolsep{\fill}}l|lllll}
    \hline
    $|\bm{k}|$=$1$ &$1$, &$\cos(\pi/5)$, &$\sin(\pi/5)$, &$\sin(2\pi/5)$, &$\cos(2\pi/5)$\\
    \hline
    $|\bm{k}|$=$q$ &$q$, &$q\cos(\pi/5)$, &$q\sin(\pi/5)$, &$q\sin(2\pi/5)$, &$q\cos(2\pi/5)$\\
    \hline
  \end{tabular*}
\end{table}

\begin{figure}[!hbpt]
	\centering
	\includegraphics[width=0.52\textwidth]{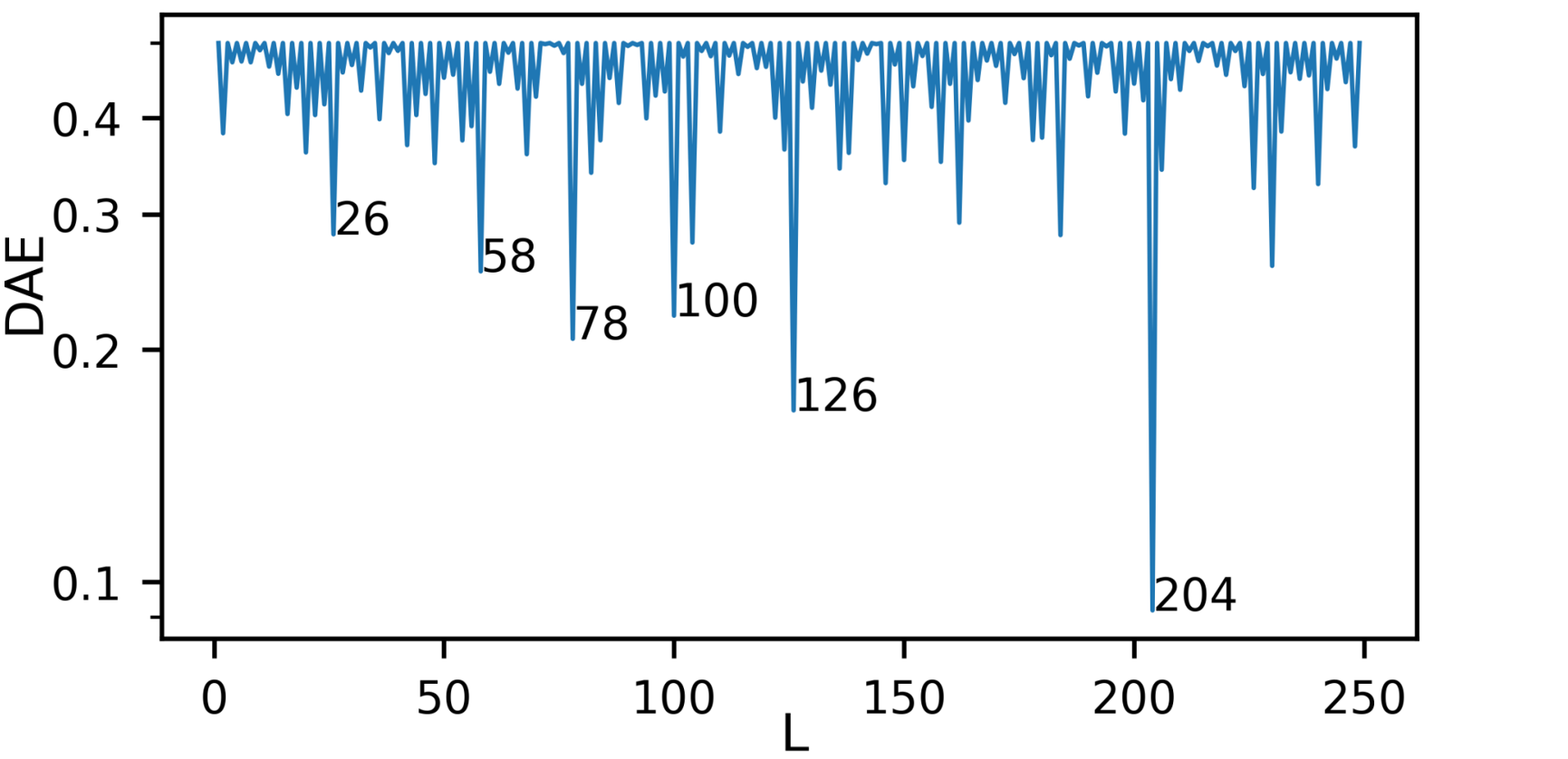}
	\caption{DAE of different computational domain sizes $[0, 2\pi L]^2$ for studying the phase transitions involving two-dimensional DQC.}
	\label{fig:Diophantine_error}
\end{figure}

FIG.\,\ref{fig:Diophantine_error} shows the DAEs as the integer $L$ increases\,\cite{davenport1946simultaneous,jiang2014numerical,yin2021transition}. In this work, we select $L=126$ ($DAE = 0.167$) and $L=204$ ($DAE = 0.092$) to obtain the proper computational domains, which can encompass the critical nucleus in the phase transition.

\subsection{Nullspace-preserving saddle search method}
Quasicrystals can be embedded into high-dimensional periodic systems and thus be viewed as having translational invariance in superspace. For the energy functional \eqref{eq:LP}, critical points (local minima/maxima, saddle points) characterizing DQC and its symmetry-related ordered structures are usually degenerate\,\cite{yin2021transition,bao2024convergence}. For a degenerate local minima $\psi_0$, its Hessian $H(\psi_0) = \nabla^2 \mathcal{F}(\psi_0)$ usually has a nullspace, whose dimension $k$ usually indicates that the corresponding ordered structure is periodic in $k$-dimensional space\,\cite{cui2024efficient, bao2024convergence}. 
The presence of the nullspace may 
prevent escape from the attraction basin of $\psi_0$ because the eigenvectors corresponding to zero eigenvalues would be mistaken as the ascent direction.

The NPSS method is an efficient saddle search method for ordered phase transitions, which can escape from the basin quickly by exploiting the properties of nullspaces and symmetry-breaking\,\cite{cui2024efficient}. In particular, at a degenerate state $\psi$ in the attraction basin of $\psi_0$, keeping the ascent direction $v$ orthogonal to the nullspace $\mathcal{W}^k(\psi)$ can eliminate the effects of nullspace. To avoid computing the nullspace at each iteration, the NPSS method uses the nullspace $\mathcal{W}^k(\bar{\psi})$ of the initial state $\bar{\psi} = \psi_0$ to replace that of $\psi$. Then the NPSS keeps the ascent direction $v$ orthogonal to the nullspace $\mathcal{W}^k(\bar{\psi})$. When the difference between nullspaces of the current and initial states becomes distinct, we update $\bar{\psi} = \psi$ 
and continue to climb as above. These operations ensure the effectiveness of the ascent direction and reduce the costs of updating the nullspaces. Therefore, the NPSS method updates the state $\psi$ by
	\begin{align}
		\beta^{-1}\dot{\psi}= -\mathcal{P}_{\mathcal{V}}T(\psi) +(I- \mathcal{P}_{\mathcal{V}})T(\psi),
		\label{eq:update-u}
	\end{align}
where $T(\psi) = -\nabla \mathcal{F}(\psi)$ is the negative gradient and $\beta$ represents the positive relaxation constant. $\mathcal{P}_{\mathcal{V}}$ denotes the orthogonal projection operator onto the subspace $\mathcal{V}$. Furthermore, the ascent direction $v$ can be updated by
\begin{align}
    \xi^{-1}\dot{v} & =-H(\psi) v+\left\langle v, H(\psi) v\right\rangle v + 2\sum_{i=1}^{l_k}\langle \bar{v}^{k}_{i}, H(\psi) v \rangle  \bar{v}^{k}_{i} ,
    \label{eq:update-v}
\end{align}
where $\xi>0$ are relaxation parameters, $v$ satisfies the unitization constraint, and $  \{\bar{v}^{k}_{i}\}_{i=1}^{l_k}$ are basic vectors of $\mathcal{W}^k(\bar{\psi})$. 

As $\psi$ climbs upward on the potential energy surface, updating $v$ is no longer affected by the nullspace after the smallest eigenvalue of $H(\psi)$ becomes negative. Thus, we update $v$ by
\begin{align}
	\xi^{-1}\dot{v}&=-H(\psi) v +\left\langle v, H(\psi) v \right\rangle v ,
	\label{eq:update-v2}
\end{align}
This NPSS method has been shown its superiority in studying the phase transition of crystals and dodecagonal quasicrystals with an economical computational cost\,\cite{cui2024efficient}.
Numerical details can be found in previous work\,\cite{cui2024efficient}.

\section{Results and discussion}
\subsection{Nucleation of DQC from liquid}\label{nucleation_from_liquid}
\begin{figure}[!htbp]
    \begin{minipage}[t]{1\linewidth}
    \centering
    \includegraphics[width=1\textwidth]{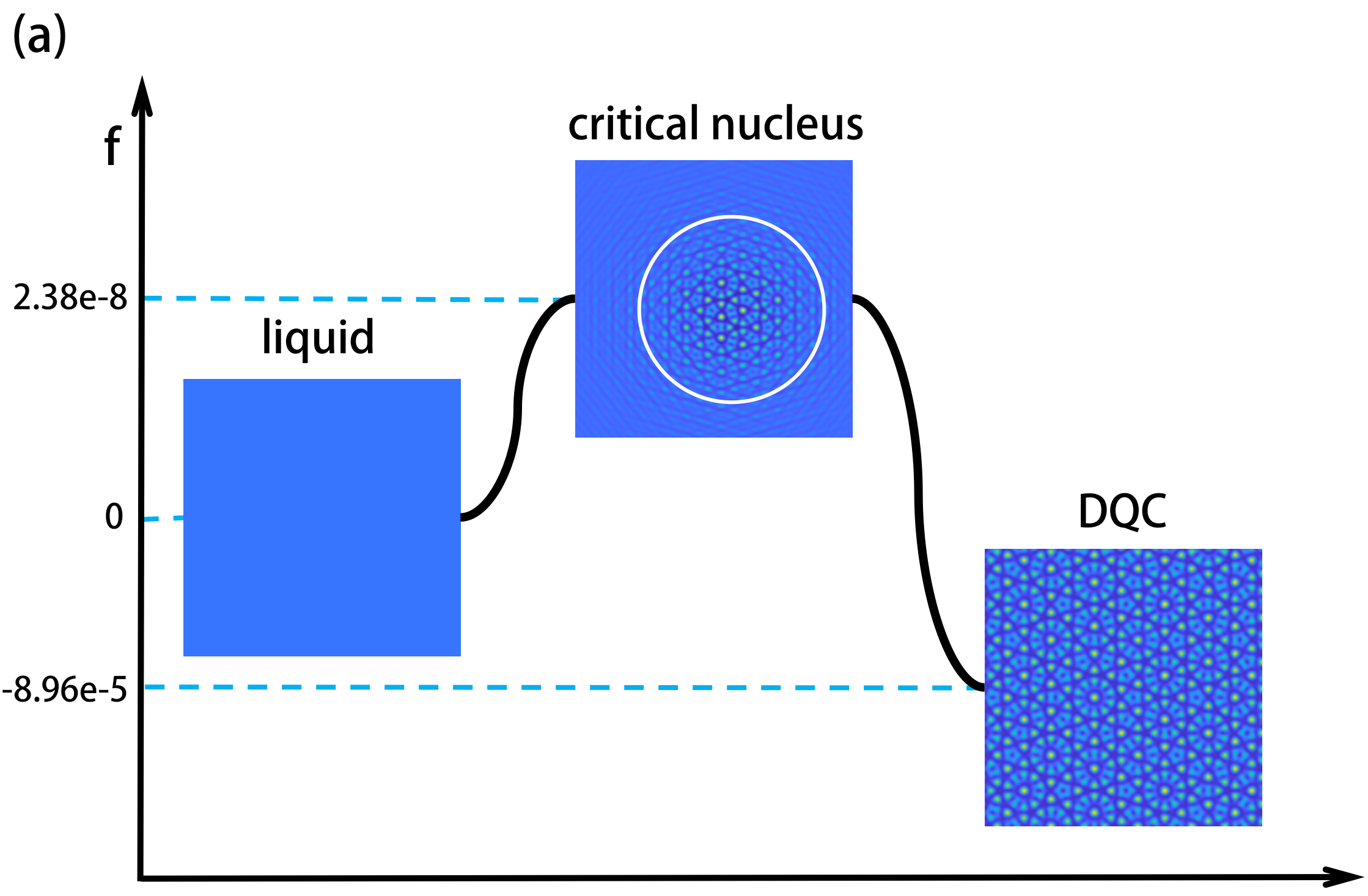}\\
    \end{minipage}%
    \\
   \begin{minipage}[t]{1\linewidth}
    \centering
    \includegraphics[width=1\textwidth]{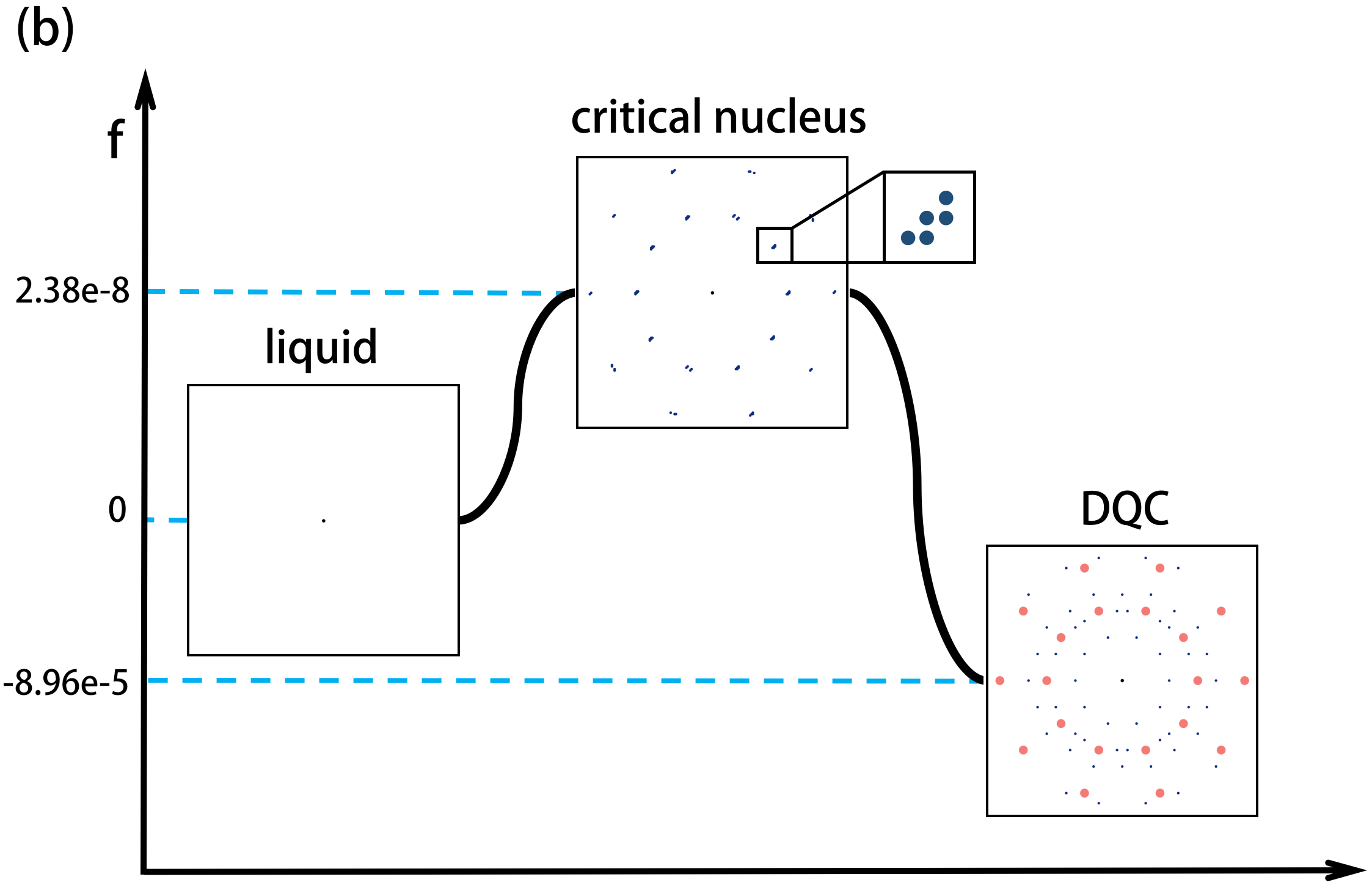}\\
    \end{minipage}%
    \caption{(a). Transition path from liquid to DQC computed by the NPSS method in the LP model with $\tau = -0.01$, $\gamma = 0.5$, where $L = 126$, $N = 1024$. (b). Diffraction spectra of stationary points on the transition path from liquid to DQC as shown in (a). The major diffraction spectra are labeled in salmon color, and the minor diffraction spectra are labeled in blue.}
    \label{fig:liquidtoDQC}
\end{figure}

In this subsection, we investigate the nucleation of DQC from the liquid state. We select an appropriate computational domain $\Omega = [0, 2\pi L]^2$ with $L = 126$ and spatial discretization points $N =1024$ in each dimension. For $\tau = -0.01$ and $\gamma = 0.5$ in the LP model \eqref{eq:LP}, DQC is the stable state with a free energy density of $f = \mathcal{F}/|\Omega|= -8.83\times 10^{-5}$, while the liquid is metastable with $f = 0$. As shown in FIG.\,\ref{fig:liquidtoDQC}\,(a), we obtain the critical nucleus of DQC with $f= 2.38 \times 10^{-8}$ by the NPSS method, which represents the transition state on the MEP. 

On the MEP, the liquid is an isotropic high-symmetry phase whose all symmetry operations belong to the extended Euclidean space group $E(\mathbb{R}^2)$\,\cite{gernoth2001analytical}. DQC has no translational invariance but a 10-fold rotation symmetry and two mutually exclusive mirror symmetries. Thus DQC is a low-symmetry phase with space group $p10mm$ in comparison to the liquid\,\cite{liu2017necessary}. Obviously, $p10mm$ is a subgroup of $E(\mathbb{R}^2)$ and a group-subgroup phase transition from the high-symmetry phase to the low-symmetry phase occurs by symmetry breaking.  
From the diffraction patterns in FIG.\,\ref{fig:liquidtoDQC}\,(b), the diffraction spectra cluster of the critical nucleus starts to show an incomplete 10-fold symmetry, indicating that the symmetry of the original structure is broken and a new symmetric structure is forming. Due to the presence of energy barriers, the appearance and initial growth of the nucleus is thermodynamically unfavorable. When the size of the DQC nucleus is smaller than that of the critical nucleus, the nucleus tends to disappear. After the DQC nucleus attains a critical size, its growth becomes thermodynamically favorable, and the nucleus will grow irreversibly, thus completely transforming into the DQC.

\subsection{Phase transition from DQC to FC}\label{quasicrystal_to_crystal}
\begin{figure*}[!htbp]
    \begin{minipage}[t]{1\linewidth}
    \centering
    \includegraphics[width=1\textwidth]{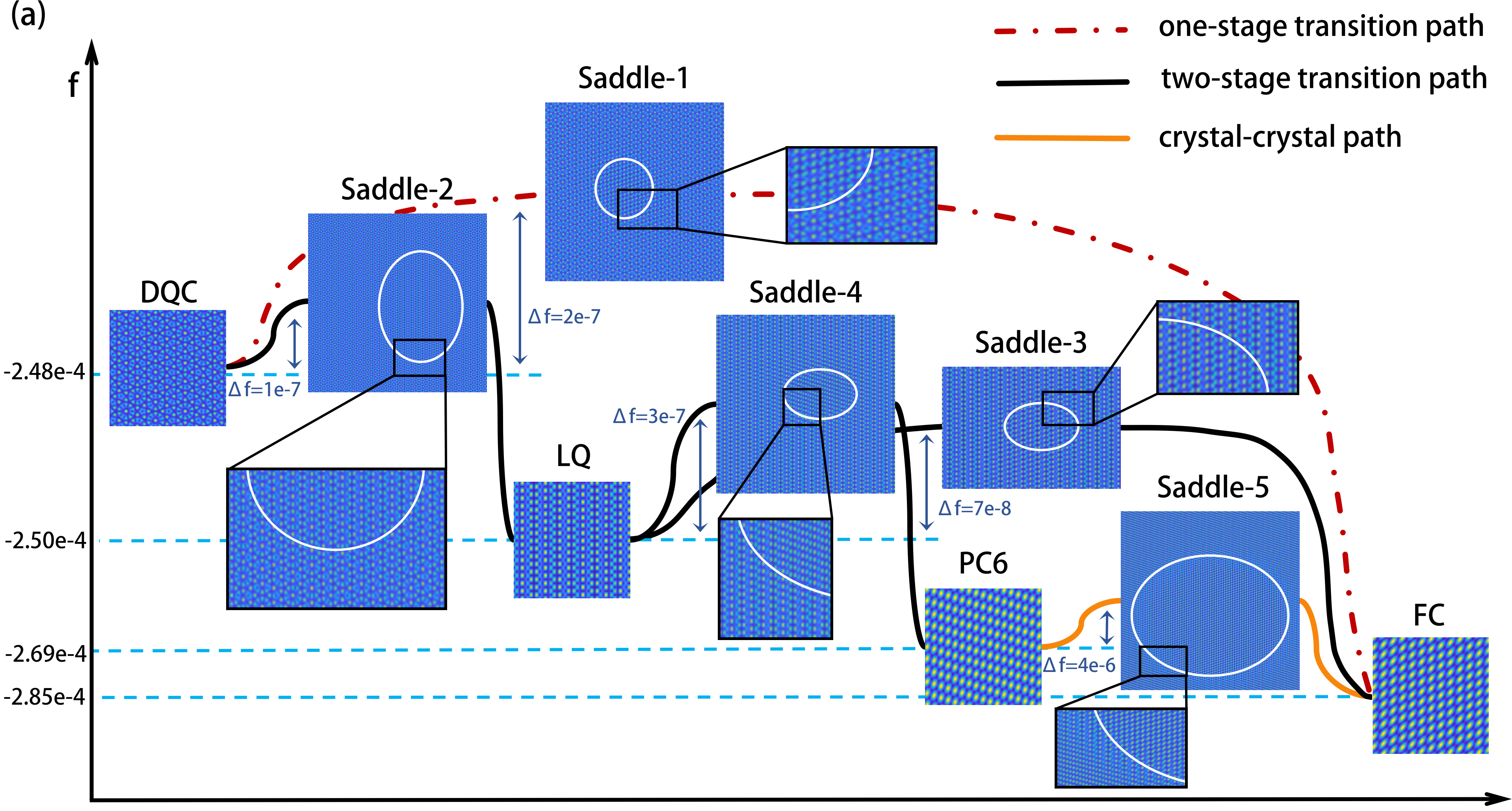}\\
    \end{minipage}%
    \\
   \begin{minipage}[t]{1\linewidth}
    \centering
    \includegraphics[width=1\textwidth]{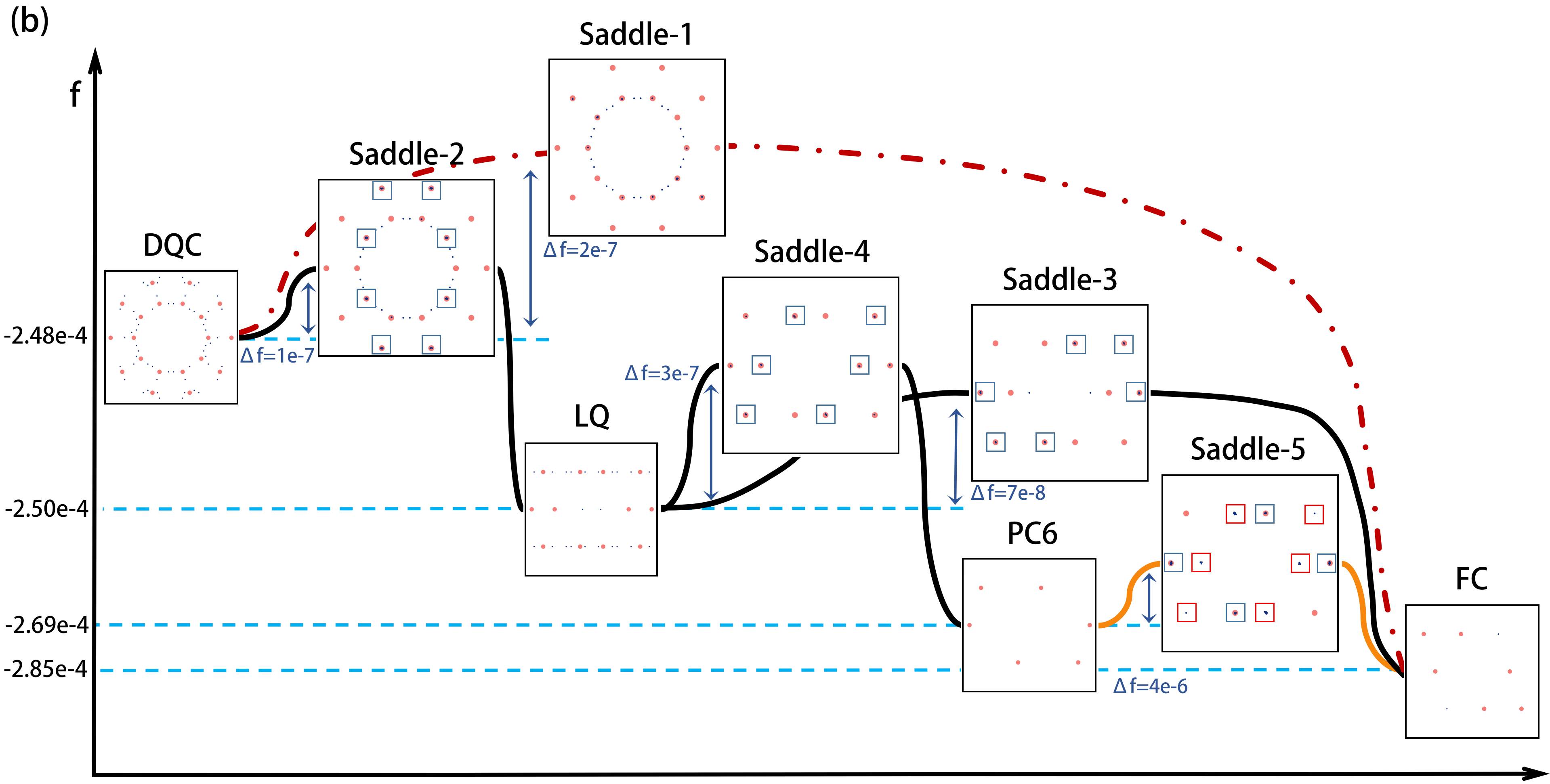}\\
    \end{minipage}%
    \caption{(a). Transition path about DQC computed by the NPSS method in LP model with $\tau= 0.018$, $\gamma = 0.4$, where $L = 126$, $N = 1024$. Saddle-i (i = 1, 2, ..., 5) are transition states on MEPs. (b). Diffraction spectra of stationary points on the transition path from DQC to FC as shown in (a). The blue and red boxes mark the diffraction spectra changed in symmetry breaking and recovery, respectively.}
    \label{fig:DQCtoFC}
\end{figure*}

We study how FC emerges from DQC. At $\tau=
0.018$, $\gamma = 0.4$ in the LP model, we obtain several paths for the phase transition from DQC to FC through the NPSS method. As shown in FIG.\,\ref{fig:DQCtoFC}, FC with $f = -2.85 \times 10^{-4}$ is more stable than the metastable state DQC with $f = -2.48 \times 10^{-4}$. The space group of FC is $p2$, which is a subgroup of $p10mm$ corresponding to DQC\,\cite{vainshtein2013fundamentals}. We observe a one-stage transition path from DQC to FC through Saddle-1 with an energy barrier $\Delta f = 2\times 10^{-7}$. This transition follows a one-step symmetry breaking mode ($p10mm \rightarrow p2$) by breaking both rotational and mirror symmetries. We then identify a metastable intermediate state LQ with the space group $p2mm$ between DQC and FC. We discover a two-stage transition path DQC $\rightarrow$ LQ $\rightarrow$ FC via the Saddle-2 and Saddle-4. The nucleation at the first stage transition from DQC to LQ shows an ellipsoidal critical nucleus of LQ with $\Delta f = 1 \times  10^{-7}$. From FIG.\,\ref{fig:DQCtoFC}\,(b), the group-subgroup phase transition from DQC reserves mirror symmetry and breaks 10-fold rotational symmetry. Consequently, LQ has a 2-fold rotational symmetry and two mutually perpendicular mirror symmetries along with horizontal and vertical axes. The second stage transition from LQ to FC involves the formation of another ellipsoidal critical nucleus of FC with $\Delta f = 3 \times  10^{-8}$. Since two mirror symmetries along the horizontal and vertical axes are broken, the space group eventually becomes $p2$. Compared to the one-stage transition, the two-stage transition with a stepwise symmetry-breaking mode ($p10mm \rightarrow p2mm \rightarrow p2$) needs to cross a lower energy barrier. It also has a higher probability of occurrence due to fewer symmetry variations at each step.

We also observe another low-symmetry phase PC6 with the space group $p2$. PC6 also has 2-fold rotational symmetry, but its lattice size with the edge length ratio $1:2\cos(\pi/5)$ is different from that of FC with the ratio $1:1$. Then we discover another two-stage phase transition path DQC $\rightarrow$ LQ $\rightarrow$ PC6, which also satisfies the stepwise symmetry-breaking mode ($p10mm \rightarrow p2mm \rightarrow p2$). By observing the diffraction spectra in FIG.\,\ref{fig:DQCtoFC}\,(b), we can find that FC and PC6 are transformed from LQ via two different symmetry-breaking modes, respectively. In particular, these two low-symmetry phases PC6 and FC are also connected by a special crystal-crystal phase transition. Firstly, the symmetries are recovered to incomplete $p2mm$, as depicted by the diffraction spectrum clusters at Saddle-5 in FIG.\,\ref{fig:DQCtoFC}\,(b). Then the recovered symmetries are broken to form FC. In this process, a higher energy barrier $\Delta f = 4\times 10^{-6}$ exists compared to other paths only with symmetry breaking. Possible causes include a high change in overall symmetry, as well as the simultaneous occurrence of symmetry breaking and recovery.

Furthermore, we analyse the nucleation rate $J$ by the formula\,\eqref{eq:nucleation rate}\,\cite{heo2010incorporating,kelton2010nucleation,ruckenstein2016kinetic,li2013nucleation,li2012numerical},
\begin{align}
  J = J_0\exp(-\Delta f),
  \label{eq:nucleation rate}
\end{align}
where $\Delta f$ represents the energy barrier and $J_0 > 0$ is the kinetic prefactor. The value of $J_0$ does not influence the criterion of
nucleation rates. From \eqref{eq:nucleation rate}, it is easy to find that the nucleation rates from DQC to Saddle-2 and LQ to Saddle-3 are higher. This also indicates that the transition with a stepwise symmetry breaking is more likely than the transition with a one-step symmetry breaking.


\subsection{Phase transition from FC to DQC}\label{crystal_to_quasicrystal}
\begin{figure*}[!ht]
    \begin{minipage}[t]{1\linewidth}
    \centering
    \includegraphics[width=1\textwidth]{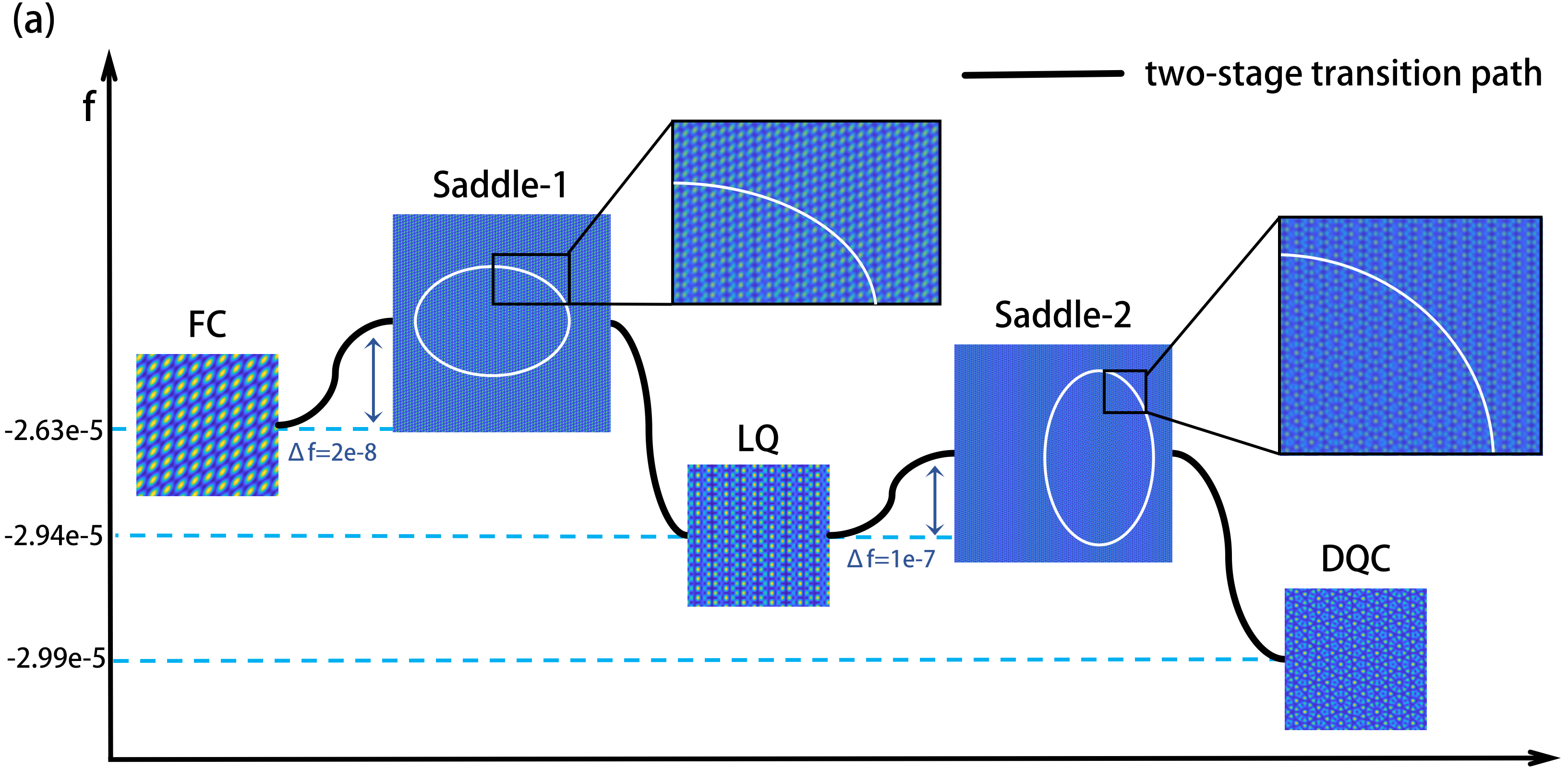}\\
    \end{minipage}%
    \\
   \begin{minipage}[t]{1\linewidth}
    \centering
    \includegraphics[width=1\textwidth]{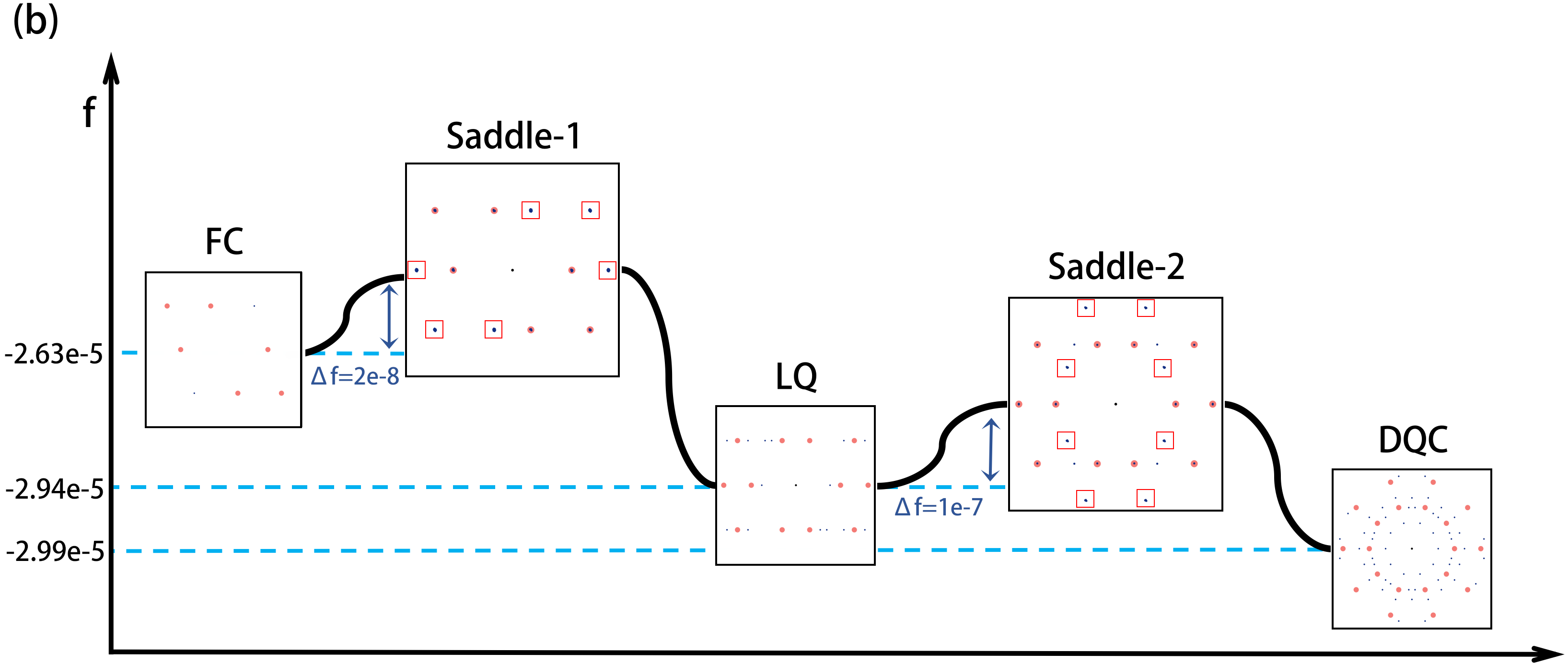}\\
    \end{minipage}%
    \caption{(a). Transition path from FC to DQC computed by the NPSS method in LP model with $\tau = -1\times 10^{-4}$, $\gamma = 0.32$, where $L = 204$, $N = 2048$. Saddle-i (i = 1, 2) are transition states on these transition paths. (b). Diffraction spectra of stationary points on the transition path from FC to DQC as shown in (a). The marked boxes have the same meaning as FIG.\,\ref{fig:DQCtoFC}.}
    \label{fig:FCtoDQC}
\end{figure*}

 In this subsection, We explore the stepwise emergence of DQC from FC. By selecting $\tau = -1\times 10^{-4}$ and $\gamma  = 0.32$ in the LP model, we observe that FC becomes a metastable state with $f=-2.63\times 10^{-5}$ and DQC reaches a stable state with $f = -2.99 \times 10^{-5}$. We discover a two-stage transition path  FC $\rightarrow$ LQ $\rightarrow$ DQC via a metastable intermediate state, as shown in FIG.\,\ref{fig:FCtoDQC}. Here, FC is a low-symmetry phase whose space group is a subgroup of that of DQC. The group-subgroup phase transition occurs through stepwise symmetry recovery. In the first stage, the mirror symmetries recover from $p2$ to $p2mm$ via an ellipsoidal critical nucleus of LQ, which is quasiperiodic in the major axis direction. In the second stage, the space group becomes $p10mm$ with rotational symmetry recovery, and a quasiperiodic order is formed in the remaining periodic direction. As shown in FIG.\,\ref{fig:FCtoDQC}\,(b), comparing the recovered diffraction spectra 
in Saddle-1 and Saddle-2, a fact is revealed that a larger symmetry variation corresponds to a higher energy barrier. However, the one-stage transition from FC to DQC is not observed, perhaps because the two-stage transition path with stepwise symmetry recovery is the more likely path. Another possible reason is that the attraction basins of FC and DQC on the potential energy surface are not adjacent. Alternatively, there may be an LQ attraction basin between them. Once the system escapes the FC attraction basin, it will easily fall into the LQ attraction basin.


\section{Conclusions}
In summary, we employ an efficient order-order phase transition algorithm to obtain the nucleation path of DQC from the liquid, and the one- and two-stage transition paths between DQC and crystals. We provide a perspective of subgroup-group phase transitions and nucleation rates to further elucidate the transition mechanisms of DQC. The results reveal that the phase transitions from DQC could follow one-step and stepwise symmetry-breaking modes to two distinct low-symmetry phases, FC and PC6. Interestingly, these two low-symmetry phases can also be linked by undergoing transitions with simultaneous symmetry recovery and breaking. Importantly, a larger symmetry variation necessitates overcoming higher energy barriers. Meanwhile, the phase transitions with the stepwise mode of symmetry breaking or recovery are more likely to occur. Overall, our study offers a comprehensive understanding of the nucleation mechanisms and phase transition pathways of DQC.

\nocite{*}
\bibliography{reference}

\end{document}